\documentclass{article}
\usepackage{spconf}

\usepackage{amsmath, bbm}
\usepackage{amssymb}
\usepackage{amsfonts}
\usepackage{amsthm}
\usepackage{algorithm}
\usepackage{algorithmic}
\usepackage{nicefrac}
\usepackage{bm}
\usepackage{hyperref}
\usepackage{cite}
\usepackage[utf8]{inputenc} 
\usepackage[T1]{fontenc}
\usepackage{graphicx}
\usepackage[font=small]{caption}
\usepackage{tikz}
\usetikzlibrary{positioning}
\usepackage{pgfplots}
\usepackage{epstopdf}
\usepackage[labelformat=simple]{subcaption}
\usepackage{multirow}
\usepackage{lipsum}

\newtheorem{Problem}{Problem}

\pgfplotsset{compat=1.16}

\input{mysymbol.sty}

\title{Distributed Scheduling using Graph Neural Networks}

\name{Zhongyuan Zhao$^\star$, Gunjan Verma$^\dag$, Chirag Rao$^\dag$, Ananthram Swami$^\dag$, and Santiago Segarra$^\star$
\thanks{Research was sponsored by the Army Research Office and was accomplished under Cooperative Agreement Number W911NF-19-2-0269. 
		The views and conclusions contained in this document are those of the authors and should not be interpreted as representing the official policies, either expressed or implied, of the Army Research Office or the U.S. Government. 
		The U.S. Government is authorized to reproduce and distribute reprints for Government purposes notwithstanding any copyright notation herein.
		\newline
		Emails:  \{zhongyuan.zhao, segarra\}@rice.edu, \{gunjan.verma.civ, chirag.r.rao.civ, ananthram.swami.civ\}@mail.mil. }}
\address{$^\star$Rice University, USA  \hspace{1cm} $^\dag$US Army’s CCDC Army Research Laboratory, USA}

\begin{document}
\ninept
\renewcommand{\baselinestretch}{0.95}
\maketitle
\begin{abstract}
A fundamental problem in the design of wireless networks is to efficiently schedule transmission in a distributed manner. 
The main challenge stems from the fact that optimal link scheduling involves solving a maximum weighted independent set (MWIS) problem, which is NP-hard. 
For practical link scheduling schemes, distributed greedy approaches are commonly used to approximate the solution of the MWIS problem.
However, these greedy schemes mostly ignore important topological information of the wireless networks.
To overcome this limitation, we propose a distributed MWIS solver based on graph convolutional networks (GCNs). 
In a nutshell, a trainable GCN module learns topology-aware node embeddings that are combined with the network weights before calling a greedy solver.
In small- to middle-sized wireless networks with tens of links, even a shallow GCN-based MWIS scheduler can leverage the topological information of the graph to reduce in half the suboptimality gap of the distributed greedy solver with good generalizability across graphs and minimal increase in complexity.
\end{abstract}
\section{Introduction}\label{sec:intro}

Wireless multi-hop networks are widely used in military and civilian applications including battlefield communications, vehicular/flying ad-hoc networks, and wireless sensor networks \cite{Lin06,sarkar2013ad}.
A fundamental problem in managing wireless multi-hop networks is to efficiently schedule the transmissions in a distributed manner. 
In general, the scheduling problem involves determining which links should transmit and when should they transmit, along with other relevant parameters such as transmit power, modulation, and coding schemes~\cite{Joo09,marques2011optimal}. 
In this paper, we focus on link scheduling in wireless networks with {time-slotted} orthogonal access, a modality often preferred over {packet-based multiple access} (e.g. CSMA-CA) due to its increased spectrum utilization efficiency~\cite{Jindal13}.
This benefit, however, comes with an associated challenge, namely that optimal scheduling involves solving a maximum weighted independent set (MWIS) problem~\cite{basagni2001finding,Joo09,joo2015local,joo2015distributed,marques2011optimal,sanghavi2009message,du2016new,Li18Ising,Douik2018,paschalidis2015message,joo2010complexity}, which is NP-hard~\cite{joo2010complexity}. 
In this context, researchers and practitioners have resorted to efficient approximation heuristics, such as message passing~\cite{sanghavi2009message,paschalidis2015message,du2016new,Li18Ising,Douik2018} and greedy solvers~\cite{Joo09,joo2015distributed,joo2015local}.

There are two main aspects central to the design of a distributed link scheduling scheme: 
1) A per-link utility function to compute the weights associated with activating each link, and 
2) An efficient distributed (approximate) solver for the associated MWIS problem. 
In terms of the per-link utility design, the queue length~\cite{Joo09,joo2015local}, link rate~\cite{Douik2018}, the product of these two~\cite{joo2015distributed} and their ratio~\cite{paschalidis2015message} have been used in the past, along with some more theoretically-grounded variations~\cite{marques2011optimal}.
However, the focus of the current paper is on the second step, i.e., the MWIS solver and, as such, we are agnostic to the specific per-link utility function selected.
The classical family of distributed heuristics for the solution of the MWIS problem can be categorized into message passing~\cite{sanghavi2009message,paschalidis2015message,du2016new,Li18Ising,Douik2018} and distributed greedy~\cite{Joo09,joo2015distributed,joo2015local} algorithms. 
This second class of algorithms is of special interest to us, where the distributed schemes seek to mimic an iterative solution obtained by adding the link with the largest utility to the partial solution and removing other links that might interfere with it.
Due to its low communication and computational complexity, variants of the distributed greedy algorithm are the common choice of practical scheduling schemes~\cite{Joo09,joo2015distributed,joo2015local}. 

Motivated by the success of machine learning in other fields, data-driven learning-based solutions for resource allocation problems in wireless networks have been proposed over the last years~\cite{lee2018deep, wang2018deep, nasir2019multi, zhang2019deep, qin2019deep,chowdhury2020unfolding}.
A common practice in most of the above frameworks is to parameterize a function of interest using multi-layer perceptrons or convolutional neural networks (CNNs), which are not well-suited for problems in wireless communications since they do not exploit the underlying topology.
This has led to several approaches that tried to adjust CNNs to the wireless setting \cite{lee2018deep,xu2019energy,van2019sum,cui2019spatial}. 
Here, we adopt an alternative direction~\cite{eisen2020optimal,chowdhury2020unfolding}, where graph neural networks (GNNs) \cite{kipf2016semi,gama2018convolutional,roddenberry_2019_hodgenet,yang_2018_enhancing} are used to incorporate the topology of the wireless network in the learning algorithm.
Our approach is also in line with the recent trend of using deep learning to find approximate solutions to combinatorial problems on graphs~\cite{khalil2017learning,li2018combinatorial}.
More precisely, we propose a modular structure where a graph convolutional network (GCN)-based node embedding module is followed by a distributed greedy algorithm, thus exploiting the efficiency of the latter while raising graph-awareness through the use of the GCN.
The GCN can be trained in an unsupervised manner -- i.e., without the need to exactly solve any MWIS problem -- and, although the training must be centralized, the deployment is fully distributed.

\noindent {\bf Contribution.} The contributions of this paper are twofold. First, we propose the first GCN-based distributed MWIS solver for link scheduling by combining the learning capabilities of GCNs and the efficiency of greedy MWIS solvers. Second, through numerical experiments, we demonstrate the superior performance of the proposed method as well as its generalizability over different graph types {and weight distributions}.

\section{System Model and Problem Statement}
\label{sec:problem}

\begin{figure}
	\vspace{-0.1in}
	\centering
	   \begin{subfigure}[b]{0.55\linewidth}
		\includegraphics[width=\linewidth]{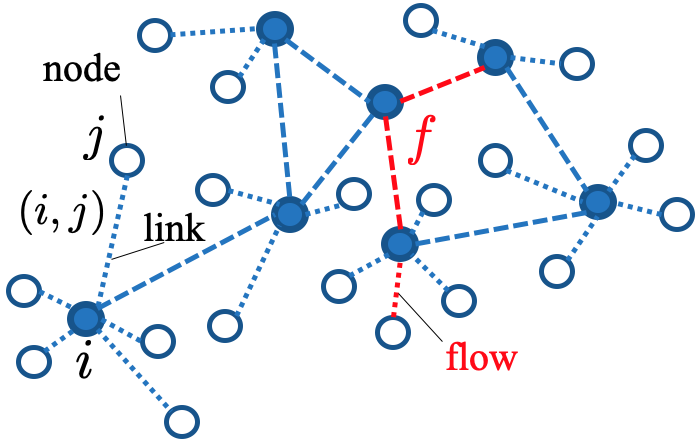}\vspace{-0.05in}
		\caption{}
		\label{fig:mhw}
	\end{subfigure}%
	~     
	\begin{subfigure}[b]{0.42\linewidth}
		\includegraphics[width=\linewidth]{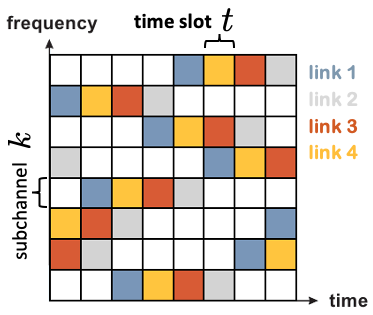}\vspace{-0.05in}
		\caption{}
		\label{fig:fdma}
	\end{subfigure}  
	\vspace{-0.1in}
	\caption{Wireless multihop network with orthogonal access. (a) Connectivity graph of the wireless multi-hop network. (b) Example of orthogonal access in an FDMA system, where the spectrum is divided into subchannels and time slots, and each spectral-temporal slot can be accessed by one in a set of potentially interfering links.} \label{fig:system}
	\vspace{-0.2in}
\end{figure}

Consider a wireless multi-hop network as illustrated in Fig.~\ref{fig:mhw}, where nodes represent users and an (undirected) link $(i,j)$ implies that user $i$ and user $j$ can communicate with each other. 
A flow $f$ describes the stream of packets from a source node to a destination node. 
A flow may pass through multiple links determined by a routing scheme. 
At each node, there is a queuing system $q$ for packets of all the flows as well as exogenous arrivals.

The interference relationship between links of the wireless multi-hop network is described by a \emph{conflict graph} $\mathcal{G}$, where a vertex in the conflict graph represents a link in the wireless network and the presence of an edge in $\ccalG$ encodes the fact that the corresponding links interfere with each other (simultaneous transmission will cause outage) or share a common node (interface constraint). 
Depending on the air-interface technology and antenna systems, the interference zone of a link can be different from its connectivity zone. 
For the rest of this paper, we focus on the conflict graph $\mathcal{G}$ which we assume to be known; see, e.g.,~\cite{yang2016learning} for its estimation. 
Notice that, in principle, the interference zone of a link (and hence $\mathcal{G}$) would depend on the transmit power of the corresponding user.
To simplify the analysis and avoid this dependence, we consider the scenario in which all the users transmit at a constant power.
Recall that an independent (vertex) set in a graph is a set of nodes such that no two nodes in the set are neighbors of each other.
From the definition of $\mathcal{G}$, only wireless links that form independent sets in $\mathcal{G}$ can communicate simultaneously in time and frequency under orthogonal access.

For multiple access, the spectrum resource is divided into $K$ orthogonal subchannels and time slots. 
The wireless channel is assumed to be stationary and ergodic, and coherent within a time slot. 
In Fig.~\ref{fig:fdma}, an example of a frequency division multiple access (FDMA) system is illustrated.\footnote{In code division multiple access (CDMA) system, channels are implemented by orthogonal codes.} 
The channel state information of link $(i,j)$ on subchannel $k$ at time slot $t$ is denoted by $h_{t}^{k}(i,j)$. 
We assume an orthogonal access scheme where each slot of the spectral-temporal grid in Fig.~\ref{fig:fdma} can only be accessed by one link out of a set of potentially interfering links.

Given the introduced system, its state at time $t$ can be described by the tuple $(\mathcal{G}_{t}^k, \mathbf{q}_t, \mathbf{f}_t, \mathbf{h}_t^k)$ consisting of the conflict graph $\mathcal{G}_{t}^k$ (potentially changing over time and subchannel), queue lengths $\mathbf{q}_t$, flows $\mathbf{f}_t$, and channel states $\mathbf{h}_t^k$.
Since the scheduling is conducted on each subchannel $k$ and at each time slot $t$, for notational simplicity we omit the subscript $t$ and superscript $k$, denoting the system state as $(\mathcal{G}, \mathbf{q}, \mathbf{f}, \mathbf{h})$. 
In this setting, the task of optimal link scheduling can be described as selecting a set $\bbv$ of links to transmit subject to orthogonality constraints and maximizing some utility $u(\bbv) = f(\bbv; \mathcal{G}, \mathbf{q}, \mathbf{f}, \mathbf{h})$ that is parametrized by the current state of the system.
As is customary \cite{marques2011optimal}, we model here the utility of the set of links $\bbv$ (or the set of nodes in the conflict graph) as the sum of utilities associated with each link $v \in \bbv$, leading to the following formal problem statement.

\begin{figure*}[t]
	\centering
	\vspace{-0.18in}
	\includegraphics[width=.85\linewidth]{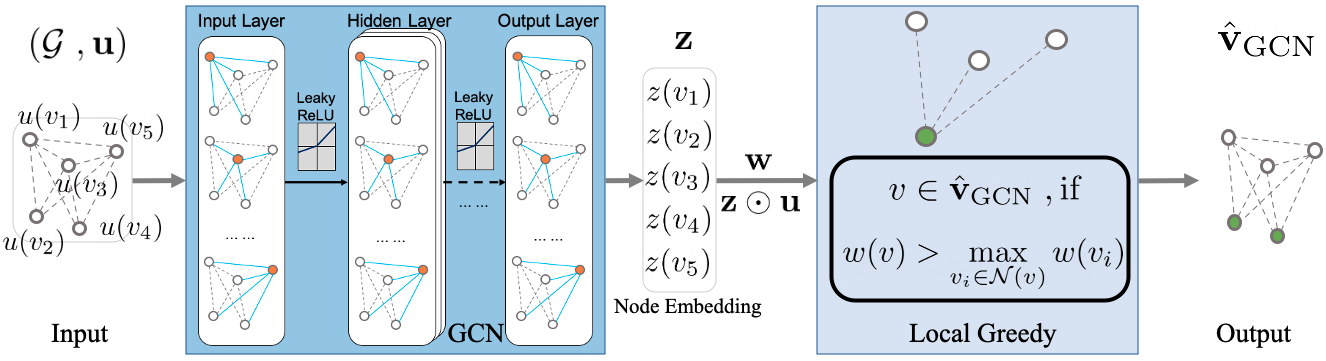}\vspace{-0.1in}
	\caption{Architecture of GCN-based distributed MWIS solver. First, the information in the conflict graph $\ccalG$ and utilities $\bbu$ is encoded into the scalar embeddings $\bbz$ via a GCN. Then, the element-wise product of $\mathbf{z}$ and $\mathbf{u}$ is fed into a local greedy algorithm to generate a solution $\hat{\bbv}_{\mathrm{GCN}}$.}
	\label{fig:dqnfw}
	\vspace{-0.2in}
\end{figure*}

\begin{Problem}\label{P:main}
    Consider a conflict graph $\ccalG$ and a utility function $u: \ccalV \to \reals_+$ where $\ccalV$ is the set of nodes in the conflict graph (links in the wireless network). The optimal scheduling is given by selecting a subset of nodes $\bbv^* \in \ccalV$ such that
		\begin{equation}\label{eq:mwis}
		\mathbf{v}^* \in \argmax_{\mathbf{v} \in \mathcal{S}(\mathcal{G})} \,\, \sum_{v\in\mathbf{v}} u(v),
	\end{equation}
	where $\mathcal{S}(\mathcal{G})$ is the set of all maximal independent sets (MIS) of $\mathcal{G}$.
\end{Problem}

From the statement of Problem~\ref{P:main}, it becomes evident that we have transformed the optimal scheduling at each spectral-temporal slot to an MWIS problem in the corresponding conflict graph. 
Indeed, we want to choose non-neighboring nodes in the conflict graph (i.e., non-interfering links in the wireless network) such that the total utility is maximized. 
As discussed in Section~\ref{sec:intro} and formally introduced here, this utility $u$ is a function of the current state of the network with many existing variants~\cite{Joo09,joo2015local,Douik2018,joo2015distributed,paschalidis2015message,marques2011optimal}.
In our case, we decide to be agnostic to the specific choice of this function and propose a solution to Problem~\ref{P:main} that is valid for any choice of utility.

\section{Distributed MWIS solver using graph neural networks}
\label{sec:solution}

A greedy approach towards approximating the solution to~\eqref{eq:mwis} would consist in building the estimate $\hat{\bbv}_{\mathrm{Gr}}$ in an iterative fashion by first adding to $\hat{\bbv}_{\mathrm{Gr}}$ the node with the largest utility, deleting its neighbors as potential candidates, and repeating this procedure until all nodes are either added to $\hat{\bbv}_{\mathrm{Gr}}$ or deleted.
By construction, we can guarantee that $\hat{\bbv}_{\mathrm{Gr}} \in \mathcal{S}(\mathcal{G})$ but the suboptimality gap $u(\bbv^*) - u(\hat{\bbv}_{\mathrm{Gr}})$ might be large since this greedy approach does not fully consider the topology of $\ccalG$.
Nonetheless, the greedy solver has two main advantages: i) Low computational complexity, and ii) Can be approximated in a distributed manner with low communication cost. 
Our goal is to \emph{retain these two salient points while decreasing the suboptimality gap}.
We propose to do this by solving~\eqref{eq:mwis} through a greedy approach but where the node weights are not given by the utilities $u(v)$ but rather by modified (graph-aware) utilities $w(v)=z(v) u(v)$, where the scalar node embedding $z(v)$ encodes a relevant topological feature of node $v$.
Intuitively, if node $v$ is a central node (a high interference link in the wireless network) we want $z(v)$ to downscale the utility $u(v)$ since scheduling $v$ would preclude us from scheduling many other links. 
By contrast, if node $v$ is peripheral in $\ccalG$ (an isolated link in the original wireless network) then $z(v)$ should amplify $u(v)$. 
In summary, $z(v)$ should be a topology-aware scaling that reduces the MWIS suboptimality gap and, to be consistent with our goal, should also be attainable in a distributed manner with low computational cost.
With these requirements in mind, we propose to obtain $\bbz$ (a vector collecting $z(v)$ for all $v \in \ccalV$) as the output of a GCN~\cite{kipf2016semi}.

We seek to obtain $\bbz$ as $\bbz = \Psi_{\ccalG}(\bbu; \ccalO)$, where $\Psi_{\ccalG}$ is an $L$-layered GCN defined on the conflict graph $\ccalG$, $\bbu$ is a vector collecting the utilities $u(v)$ for all $v \in \ccalV$, and $\ccalO$ represents the collection of trainable parameters of the GCN.
To be more precise, if we define $\bbX^0 = \bbu$, we then have that $\bbz = \Psi_{\ccalG}(\bbu; \ccalO) = \bbX^L$, where an intermediate $l$th layer of the GCN is given by
\begin{equation}\label{E:gcn}
	\mathbf{X}^{l+1} = \sigma\left(\mathbf{X}^{l}{\bbTheta}_{0}^{l}+\bbcalL \mathbf{X}^{l}{\bbTheta}_{1}^{l}\right).
\end{equation}
In~\eqref{E:gcn}, $\bbcalL$ is the normalized Laplacian of $\ccalG$, ${\bbTheta}_{0}^{l}, {\bbTheta}_{1}^{l} \in \mathbb{R}^{g_{l} \times g_{l+1}}$ are trainable parameters, and $\sigma(.)$ is the activation function. 
The activation functions of the input and hidden layers are selected as leaky ReLUs whereas a linear activation is used for the output layer. 
{The input and output are both vectors, i.e., $g_{0}=g_{L}=1$.}
Since $\bbcalL$ is a local operator on $\ccalG$, it should be noted that $z(v)$ can be computed locally at each node $v$ through $L$ communications with its neighbors. 
As an extreme case, the 1-layer GCN does not employ any non-linear activation and node embeddings $z(v)$ can be locally computed as 
\begin{equation}\label{eq:1layer}
	z(v) = u(v) \, \theta_{0} + \Bigg( u(v) - \sum_{v_i \in \mathcal{N}(v)}d_{v}^{-\frac{1}{2}}d_{v_i}^{-\frac{1}{2}} u(v_i)\Bigg)\theta_{1},
\end{equation}
where $\mathcal{N}(v)$ denotes the neighborhood of node $v$, and $d_v$ and $d_{v_i}$ represent the degrees of nodes $v$ and $v_i$, respectively.

\begin{figure*}[!th]
\centering
\vspace{-0.18in}
        \begin{subfigure}[b]{0.32\textwidth}
                \includegraphics[height=1.55in]{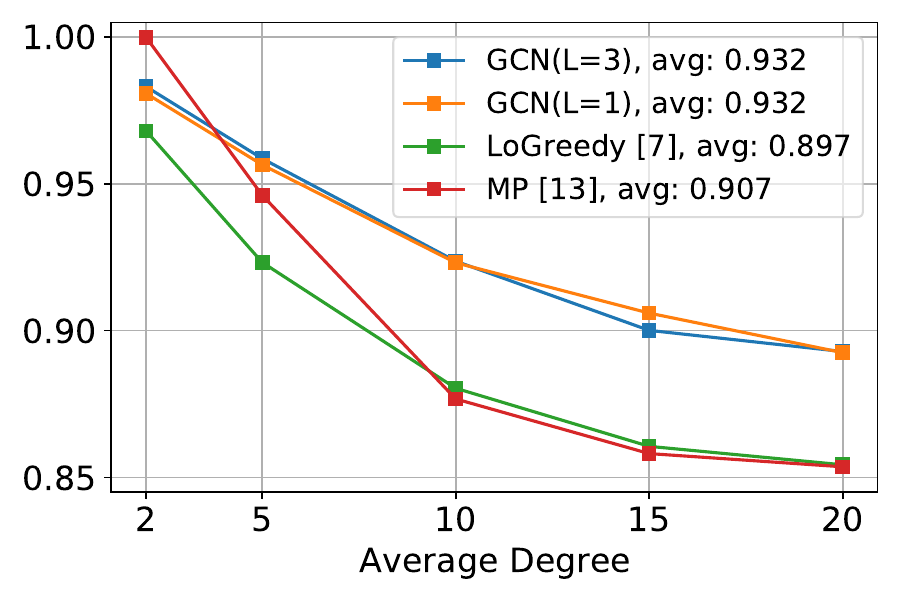}
                \vspace{-0.25in}\caption{}
                \label{fig:cmpx}
        \end{subfigure}%
~     
        \begin{subfigure}[b]{0.32\textwidth}
                \includegraphics[height=1.55in]{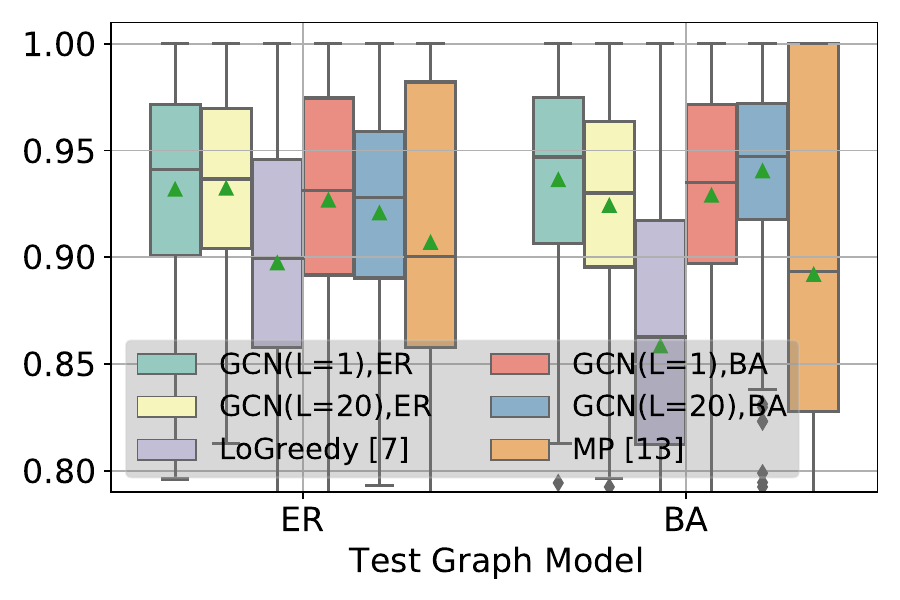}
                 \vspace{-0.25in}\caption{}
                \label{fig:generalization}
        \end{subfigure}  
~
       \begin{subfigure}[b]{0.32\textwidth}
                \includegraphics[height=1.55in]{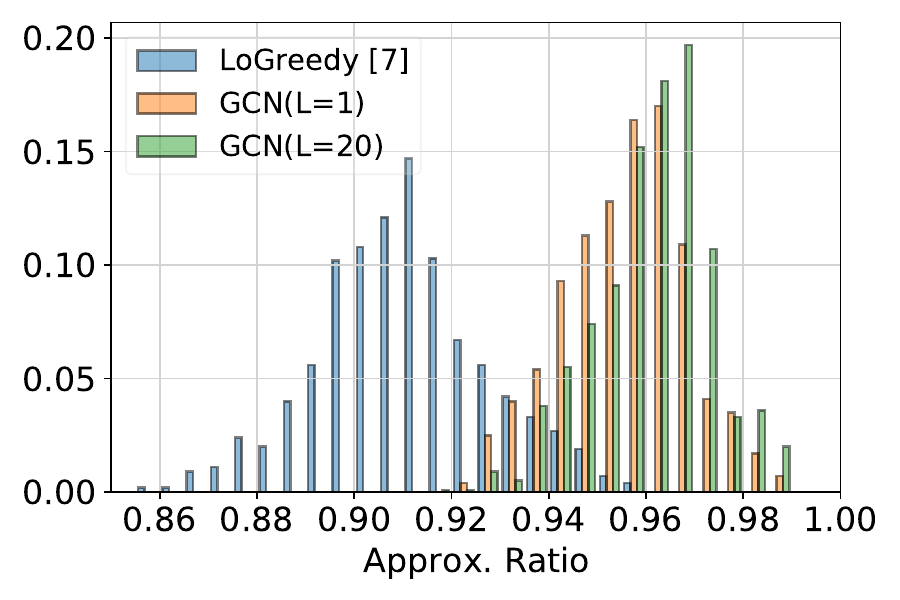}
                 \vspace{-0.25in}\caption{}
                \label{fig:traffic}
        \end{subfigure}%
        \vspace{-0.12in}
        \caption{Performance of GCN-based distributed MWIS solvers. (a) Approximation ratio as a function of the average degree on ER graphs of mixed sizes. 
        (b) Approximation ratio of GCNs trained and tested on ER and BA models. 
        (c) Histogram of approximation ratio of distributed link scheduling in wireless networks. The means of local greedy, 1-layer GCN, and 20-layer GCN are 0.911, 0.956, and 0.961,~respectively.}   
 \label{fig:label_all}    
 \vspace{-0.23in}
\end{figure*}

The architecture of the proposed distributed MWIS (approximate) solver is illustrated in Fig.~\ref{fig:dqnfw}. 
In accordance with Problem~\ref{P:main}, the inputs to our solver consist of the conflict graph $\ccalG$ and a vector $\bbu$ of node utilities, and the output of the solver is an estimate $\hat{\bbv}_{\mathrm{GCN}}$ of the optimal link scheduling.
As discussed, we first encode the graph structure in the scalar node embeddings $\bbz$ through a GCN.
We then compute the element-wise product between $\bbz$ and $\bbu$ (the original per-link utilities) to obtain a set of modified graph-aware utilities $\bbw = \bbz \odot \bbu$. 
Our estimated solution $\hat{\bbv}_{\mathrm{GCN}}$ is obtained from $\bbw$ through a local greedy algorithm~\cite{joo2015local}.
In a nutshell, this local greedy algorithm is implemented by adding vertex $v$ to the solution set as
\begin{equation}\label{eq:lg}
	v\in\hat{\bbv}\;,\text{ if }w(v) > \max_{v_i \in \mathcal{N}(v)} w(v_i)\;.
\end{equation}
Then, \eqref{eq:lg} is iterated on $\ccalG'=\ccalG \setminus (\hat{\bbv}\cup\mathcal{N}(\hat{\bbv}))$ until $\ccalG'$ is empty.
Moreover, the local greedy algorithm has a built-in tie-breaking mechanism based on an initial assignment of identification numbers to each node in the wireless network that does not require additional information exchanges in the case of a tie. 

The communication complexity (defined as the rounds of local exchanges between a node and its neighborhood) of the proposed MWIS solver is $L+\log(|V|)$, where $\log(|V|)$ is the average complexity of the local greedy algorithm \cite{joo2015local}.
In this way, $\hat{\bbv}_{\mathrm{GCN}}$ can be computed in a distributed manner, where the local computational and communication costs can be controlled by modifying the number of layers $L$ in the GCN.
Importantly, notice that the logarithmic local communication complexity is a key aspect to promote scalability.

Having discussed the rationale and the mechanics of a forward pass on our proposed architecture, we are left to discuss how to train the parameters $\ccalO$ in the GCN.
It should be noted that the proposed setting differs from the typical supervised or semi-supervised settings on which GCNs are employed in two main aspects.
First, although we can simulate training inputs ($\ccalG$, $\bbu$), in general we cannot obtain their associated optimal outputs $\bbv^*$ since this would require solving an NP-hard problem.
Second, the output $\bbz$ of the GCN is related to the objective to be maximized $u(\hat{\bbv}_{\mathrm{GCN}})$ through a non-differentiable local greedy process.
Inspired by reinforcement learning (RL) techniques, we seek to overcome these limitations by backpropagating a reward signal from $u(\hat{\bbv}_{\mathrm{GCN}})$ to $\bbz$. 
In particular, for nodes $v \in \hat{\bbv}_{\mathrm{GCN}}$, we consider the reward signal
\begin{equation}\label{E:reward}
	\rho(v) = \frac{u(\hat{\bbv}_{\mathrm{GCN}}) + u(v)}{u(\hat{\bbv}_{\mathrm{Gr}})}.
\end{equation}
Notice that this reward signal is computed based on the efficient greedy solver in~\eqref{eq:lg} (where we replace $\bbw$ with $\bbu$), thus circumventing the need to have access to exact solutions of the MWIS problem.
Intuitively, whenever the inclusion of $v$ in $\hat{\bbv}_{\mathrm{GCN}}$ leads to a solution with higher utility than the greedy baseline [$u(\hat{\bbv}_{\mathrm{GCN}}) > u(\hat{\bbv}_{\mathrm{Gr}})$] and $v$ plays a central role in this solution [large $u(v)$], we want to reward node $v$. 
By contrast, whenever node $v$ is included in a bad solution [$u(\hat{\bbv}_{\mathrm{GCN}}) < u(\hat{\bbv}_{\mathrm{Gr}})$], we want to avoid its inclusion in future solution sets.
Consequently, we propose the following root-mean-square loss to train our GCN
\begin{equation}\label{E:gcn_loss}
\vspace{-0.05in}
\ell(\ccalO; \ccalG, \bbu) = \sqrt{\sum_{v \in \ccalV} \frac{(z(v)-\rho(v))^2}{|\ccalV|}},
\end{equation}
where $|\ccalV|$ denotes the size of the graph.
By drawing graphs $\ccalG$ and utility weights $\bbu$ from prescribed distributions, we train our GCN weights $\ccalO$ through batch training, employing the Adam optimizer and exponentially decaying learning rates. 

\vspace{-0.5mm}
\section{Numerical experiments}
\label{sec:results}
\vspace{-0.5mm}

We illustrate the performance of the proposed GCN-based distributed MWIS solver in synthetic random graphs and as a scheduler in wireless networks.
As a comparative baseline, we consider the distributed approximate solvers local greedy ~\cite{joo2015local} and message passing~\cite{paschalidis2015message}. 
The quality of an approximate solution $\hat{\mathbf{v}}$ is evaluated by its approximation ratio $u(\hat{\mathbf{v}})/u(\mathbf{v}^{*})$.
The optimal solution $\mathbf{v}^*$ is obtained by solving the computationally expensive integer programming formulation of MWIS \cite{sanghavi2009message,paschalidis2015message} using the Gurobi solver~\cite{Gurobi}.

The synthetic graphs for training and testing are generated from the Erdős–Rényi (ER)~\cite{erdds1959random} and Barabási–Albert (BA)~\cite{Albert02} models. 
The ER model is completely determined by two parameters: the number of nodes $N$ and the probability of edge-appearance $p$.
The BA model is also determined by two parameters: $N$ and the number of edges that each new node forms $m$ during the preferential attachment process.
In our experiments, we set $m=Np$ so that graphs from the ER and BA have the same expected average degree.


The evaluated GCNs have different numbers of layers. 
For $L>2$, the size of every hidden layer is $g_l=32$. 
Each GCN is trained on a dataset of 5800 random graphs, including 5000 graphs of size $N\in \left\{100, 150, 200, 250, 300\right\}$ and expected average degree $Np\in \left\{2, 5, 7.5, 10, 12.5\right\}$, and 800 graphs of size $N\in\left\{30,100\right\}$ and edge probability $p\in\left\{0.1,0.2,\dots,0.9\right\}$.
Unless otherwise specified, the random graphs in the training set are drawn from an ER model and the vertex utilities are drawn following a uniform distribution $u(v)\sim\mathcal{U}(0,1)$.
The RL settings include a batch size of 200 for experience replay, 25 epochs, and periodic gradient reset.\footnote{Training typically takes 30 minutes on a workstation with a specification of 16GB memory, 8 cores and Geforce GTX 1070 GPU. The source code is published at \url{https://github.com/zhongyuanzhao/distgcn}}

\medskip\noindent{\bf MWIS solvers on random graphs.}
We consider a test set containing 500 ER graphs of size $N\in \left\{100, 150, 200, 250, 300\right\}$ and average degree $Np\in \left\{2, 5, 10, 15, 20\right\}$, where we include $20$ instances of each pair $(N, Np)$. 
The mean approximation ratio attained as a function of the average degree of the tested graphs is illustrated in Fig.~\ref{fig:cmpx}.
Overall, the GCN-based solvers yield improvements of $3.5\%$ ($0.932$ vs $0.897$) compared to the local greedy solver (LoGreedy) \cite{joo2015local}.
Moreover, this difference is more conspicuous (close to $5\%$) for the more challenging case of graphs with larger average degrees.
Indeed, a wrong greedy selection of a high-degree node can significantly affect the suboptimality gap.
The message passing solver (MP)~\cite{paschalidis2015message} with higher  communication complexity performs similarly to local greedy except on graphs with small average degree.
Lastly, it should be noted that increasing $L$ from 1 to 3 does not yield better performance. 
This can be partially attributed to the limited topological information carried by the 1-dimensional node embedding $\mathbf{z}$.
For this simple case, this indicates that the 1-layer GCN with minimal complexity is a good choice for distributed scheduling. 

We next analyze how well the proposed scheme generalizes across different graph models.
To do this, we consider two training sets containing ER and BA models, respectively, with the rest of configurations identical to the default training set. 
Each trained GCN is then tested on the ER test of the previous experiment and a BA test set with identical configuration ($m=Np$).
GCNs of 1 layer and 20 layers are evaluated to understand the impact of $L$; see Fig.~\ref{fig:generalization} for the corresponding box plot.
Given the heavy-tailed degree distribution of BA graphs~\cite{Albert02}, we can see a marked drop in the performance of the (topology-agnostic) local greedy solver {and the message passing solver}.
Indeed, although GCNs outperform local greedy in all settings, this margin is greater in BA graphs, underscoring the value of taking the topology into account.
In general, GCNs tend to perform better if they are trained on the same graph model of the test set, but this influence is small, showing good transferability of the proposed approach. 
Compared to shallower GCNs, deeper GCNs do not significantly improve the mean approximation ratios but tend to present smaller variance. Moreover, deep GCNs are more tuned to the training graph model. For example, $L=20$ attains the best performance when trained and tested in BA, but underperforms compared to $L=1$ when trained in one graph model and tested on the other. 
{Although not shown here, similar transferability holds when varying the vertex utility distribution.}

\medskip\noindent{\bf GCN-based distributed scheduling. }
Lastly, we evaluate the distributed MWIS solvers for throughput-optimal scheduling in simulated wireless networks. 
These networks consist of $100$ nodes randomly distributed in a square of area $250$. 
A link is established between two nodes if their distance is smaller than $1$, and two links interfere with each other if they have incident nodes within the distance of $4$. 
A $1$-hop flow with random direction is created on each link. 
The exogenous packets on each source node follow a Poisson arrival with a constant flooding arrival rate. 
The link rate $r(v)$, defined as the number of packets that can be transmitted through link $v$ in a time slot, follows a discrete uniform distribution $\mathcal{U}(0,100)$. 
We set the utility of each node in $\ccalG$ as the number of packets that can be transmitted through that link, i.e., $u(v)=\min(q(v),r(v))$, where $q(v)$ is the backlog of link $v$.
We consider $100$ conflict graphs, typically with $40$-$60$ links and average degree of $13.1$.
For each conflict graph, $10$ scheduling instances of $200$ time slots are executed by each tested scheduler. In each scheduling instance, the random processes of arrivals and link rates for each scheduler are identical.

On average, the throughput of the local greedy scheduler is $91.1\%$ of the optimal, and the approximation ratios for the GCN solvers are $95.6\%$ ($L=1$) and $96.1\%$ ($L=20$). 
The histogram of approximation ratios of these three schedulers, as illustrated in Fig.~\ref{fig:traffic}, shows significant and consistent improvement of GCN over greedy. 
The deeper GCN ($L=20$) also consistently outperforms the shallow one ($L=1$) by a small margin, despite their similar mean approximation ratio on the ER test set [cf. Fig.~\ref{fig:generalization}]. 
Based on these results, the 1-layer GCN is an attractive choice for practical link scheduling since it closes the gap between optimal and greedy by half at the cost of only $1$ additional local exchange.

\section{Conclusions and future work}
\label{sec:conclusions}

We presented a distributed MWIS solver based on GCNs for link scheduling in wireless networks. 
The proposed solver can be trained on simulated networks and generalizes well across different types of graphs and utility distributions. 
By combining the efficiency of local greedy schemes and the ability of GCNs to encode topological information, the proposed approach achieves superior performance over greedy baselines with minimum increase in complexity.
Moreover, our approach is agnostic to the specific per-link utility, thus, it can be used in conjunction with many existing distributed scheduling protocols. 
Current research efforts include: i) Trading off complexity and performance by incorporating higher-dimensional node embeddings, and 
ii) Considering more sophisticated utilities (and training strategies) that take into account the temporal correlation of the MWIS problems encountered when managing a wireless network.

\vfill\pagebreak



\bibliographystyle{IEEEbib}
\bibliography{strings,refs}

\end{document}